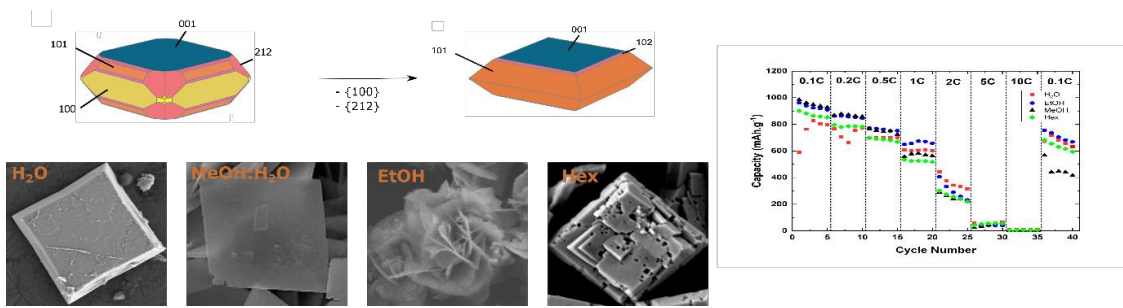

# Solvent Engineered Synthesis of Layered SnO Nanoparticles for High-Performance Anodes

**Sonia Jaśkaniec[a,b,*], Seán R. Kavanagh[b,c,d,e], João Coelho[a,b], Seán Ryan[a,b], Christopher Hobbs[b,f], Aron Walsh[d,g,h], David O. Scanlon[c,e,h,i] and Valeria Nicolosi[a,b,*]**

[a]School of Chemistry, Trinity College Dublin, Dublin 2, Ireland

[b]CRANN & AMBER, Trinity College Dublin, Dublin 2, Ireland

[c]Department of Chemistry, University College London, 20 Gordon Street, London WC1H 0AJ, UK

[d]Department of Materials, Imperial College London, Exhibition Road, London SW7 2AZ, UK

[e]Thomas Young Centre, University College London, Gower Street, London WC1E 6BT, UK

[f]School of Physics, Trinity College Dublin, Dublin 2, Ireland

[g]Department of Materials Science and Engineering, Yonsei University, Seoul 03722, Korea

[h]The Faraday Institution, Quad One, Harwell Science and Innovation Campus, Didcot, UK

[i]Diamond Light Source Ltd., Diamond House, Harwell Science and Innovation Campus, Didcot, Oxfordshire OX11 0DE, UK

metels@tcd.ie, nicolov@tcd.ie

**Abstract:** Batteries are the most abundant form of electrochemical energy storage. Lithium and sodium ion batteries account for a significant portion of the battery market, but high-performance electrochemically active materials still need to be discovered and optimized for these technologies. Recently, tin(II) oxide (SnO) has emerged as a highly-promising battery electrode. In this work, we present a facile synthesis method to produce SnO nanoparticles whose size and shape can be tailored by changing the solvent nature. We study the complex relationship between wet chemistry synthesis conditions and resulting layered nanoparticle morphology. Furthermore, high-level electronic structure theory, including dispersion corrections to account for van der Waals forces, are employed to enhance our understanding of the underlying chemical mechanisms. The electronic vacuum alignment and surface energies are determined, allowing the prediction of the thermodynamically-favoured crystal shape (Wulff construction) and surface-weighted work function. Finally, the synthesized nanomaterials were tested as Li-ion battery anodes, demonstrating significantly enhanced electrochemical performance for morphologies obtained from specific synthesis conditions.

1. Introduction

In the last decade, two-dimensional (2D) metal oxide nanoparticles have attracted much attention due to their enhanced properties in comparison to their bulk counterparts.[1-4] One such chemical compound is tin monoxide (SnO), whose layered crystal structure renders it amenable to the fabrication of 2D architectures. Moreover, it is environmentally friendly, abundant and naturally occurring as the mineral Romarchite. In its bulk form, SnO adopts a PbO-type tetragonal crystal structure with a layered -Sn-O-Sn- pattern along the *c* direction.[5] A strong driving force behind the stability of this distorted structure is an asymmetric electron distribution about the Sn atoms, resulting in a stereochemically-active $5s^2$ 'lone pair' directed toward the interlayer gap. Density Functional Theory (DFT) calculations have revealed a coupling between unfilled Sn $5p$ and an anti-bonding combination of Sn $5s$ – O $2p_z$ as the origin of this asymmetric electron density.[6]

The resulting layered structure with its sizeable interlayer spacing ($c$ = 4.84 Å) relative to the active ion size (Li atomic radius of 1.845 Å, Na atomic radius of 1.80 Å), provides fast diffusion channels, thereby offering greater intercalation of the active ions in Li/Na ion batteries and, consequently, rapid charge-discharge capability.[7,8] In addition, this structure can as well moderate volume changes during battery charge-discharge processes. With a high theoretical capacity of 875 mA.h.g$^{-1}$ and 1150 mA.h.g$^{-1}$ for lithium and sodium, respectively, SnO is a promising candidate for energy storage applications.[5,9] Energy storage capabilities of battery materials can be further improved via (i) use of nano-scale materials and (ii) subsequent tailoring of nanoparticle size and shape.[5,9-11] Various SnO morphologies, such as rose-like particles,[12,13] platelets[5,14] and other hierarchical architectures,[11,15,16] have been reported, but the exact relationship between synthesis conditions and morphology is still not fully understood. One of the reported parameters strongly affecting nanoparticles morphology is the nature of the solvents used in the synthesis. In spite of several theories describing the relationship between particles morphology and solvent properties,[17-19] this seems to be strongly system-dependent, requiring specific investigation for relevant materials. Nevertheless, for oxide particles produced by hydroxide decomposition, solvent polarity seems to play a key role in nucleation and crystal growth,[11,20,21] likely a result of control over hydrolysis of the intermediate product. For tin oxides (SnO and $SnO_2$), alcohols and their mixtures with water are widely used solvents,[11,15,18,19,21] hence we performed a comprehensive study of synthesis-morphology relationship using these solvents.

In this paper, we present size and shape-controlled synthesis of SnO layered nanoparticles obtained by wet-chemistry using various alcohols and their mixtures with water as reaction media. We analyze the influence of carbon-chain length, reaction temperature and polarity on nanoparticle morphology and size. Furthermore, first-principles electronic structure theory was employed to augment our understanding of the underlying chemical mechanisms governing nanoparticle morphology, including the prediction of crystal shape (Wulff construction) and electrochemical work function. Finally, the synthesized nanomaterials of varying morphology were tested as battery anodes.

## 2. Results and discussion

SnO nanoparticles were produced in a two-step process as detailed in the experimental section. In the first step, a white precipitate of pure tin oxide-hydroxide, $Sn_6O_4(OH)_4$, was obtained, as confirmed by X-ray diffraction (Figure 1a), JCPDS no. 46-1486. Additionally, SEM analysis (Figure 1b) of the same showed that the material consists of small nanoparticles (< 100 nm) with undefined morphology.

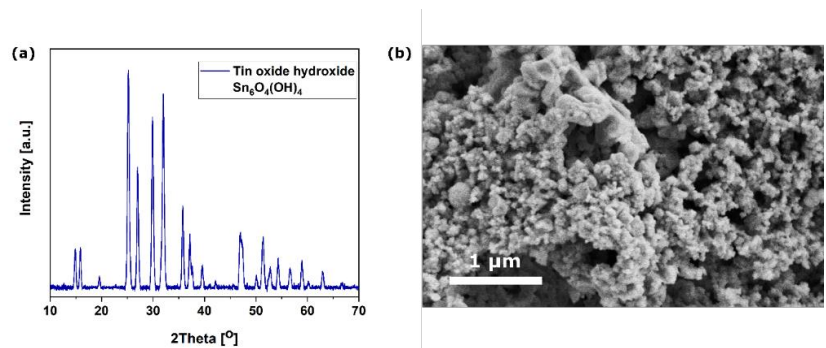

**Figure 1** a) X-ray diffraction pattern and b) SEM micrograph of tin oxide hydroxide ($Sn_6O_4(OH)_4$).

In the second synthesis step, the tin oxide-hydroxide precipitate was dispersed in different alcohols and their mixture with water. A colour change from white to brown-grey was observed after several hours of heating, which was expected for tin(II) oxide-hydroxide to tin(II) oxide transformation.[11,15]

*2.1. Synthesis of SnO in pure alcohols*

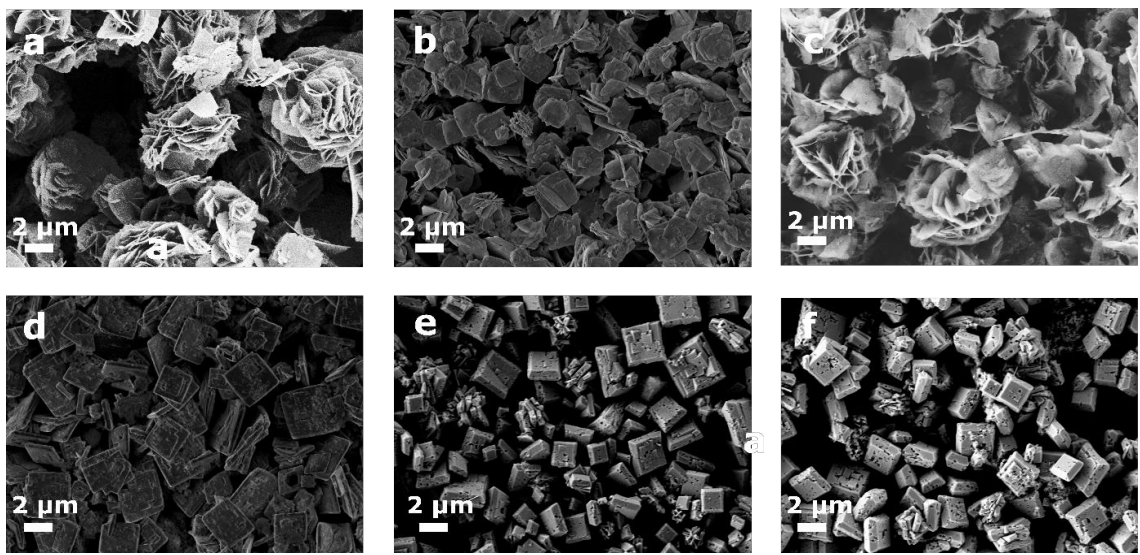

**Figure 2** SEM micrographs of SnO nanoparticles produced in refluxing a) ethanol, b) 1-propanol, c) 2-propanol, d) 1-butanol, e) 1-pentanol and f) 1-hexanol.

Firstly, tin oxide-hydroxide precipitate was refluxed in pure alcohols in the temperatures ranging from 65-157 °C. For the reaction performed in pure methanol the colour change, and the transformation from tin oxide-hydroxide was not observed, likely due to the low reaction temperature (boiling point of methanol is 65 °C),[22] and/or neutral pH.[13] For the higher boiling point alcohols, a colour change was noticed during heating. Interestingly, oxide-to-hydroxide transformation was observed only when the solvent was boiling. It was especially prominent for 1-pentanol and 1-hexanol, which boil at relatively high temperatures, as lowering the reaction temperatures only a few degrees under the boiling point did not result in hydroxide-to-oxide transformation.

For pure ethanol (boiling at 78 °C), SnO nanoparticles with well-defined, flower-like morphology were obtained, as presented in Figure 2a. Similar results were observed for the sample refluxed in pure 2-propanol (Figure 2c). Most probably, very similar properties of these two solvents (boiling point, surface tension, Hansen solubility parameters and polarity) lead to the formation of similar nanostructures. Comparable morphologies were reported for the syntheses performed at highly basic pH (it was neutral in our case) and/or hydrothermal conditions at temperatures exceeding the boiling point of the solvents (reflux at atmospheric pressure was used in our study).[9,12,13,19] SnO flowers obtained by Uchiyama et al.[13] were produced in highly basic (pH of 13.0-13.3) aqueous solution and relatively concentrated $SnF_2$ precursor (0.35-0.5 M). This very precise synthesis conditions and narrow pH range make this approach complicated and rather unsuitable for industrial-scale production, similarly to the hydrothermally treated approach used by Sun et al.[12] In contrast, the method used in our study is relatively straightforward and scalable.

The reactions performed in 1-propanol and 1-butanol resulted in the formation of smaller and thicker square-like nanoparticles (Figure 2b and d, respectively). These morphologies might be a result of higher temperatures at which the reactions were performed, which influence the hydroxide-to-oxide transformation kinetics.

SnO nanoparticles produced in 1-pentanol and 1-hexanol (Figure 2e and f, respectively) have perforated thick squares morphologies and were significantly thicker in comparison to the structures obtained in shorter carbon-chain alcohols. Thick squares were also obtained in water used as a solvent, although porous structures were not observed (Figure S1 in ESI). The formation of these nanostructures is probably a result of higher surface tension, density and dipole moment for these solvents in comparison to shorter alcohols.

The synthesis was also performed in pure ethylene glycol, in order to determine the influence of additional OH group to hydroxide-to-oxide transformation. Refluxing the tin(II) hydroxide dispersion in this solvent did not lead to the colour change. A comparable observation was reported for the synthesis of $SnO_2$,[23] where water was indispensable for the formation of the intermediate tin hydroxide, which was subsequently transformed to SnO and finally oxidized to $SnO_2$. Tin hydroxide is a starting compound in our experiments, so water-driven hydroxide formation can be excluded, but perhaps hydroxide to monoxide transformation reported in the Jiang et al.[23] is also a step requiring a suitable solvent, while ethylene glycol is only a reaction medium directing nanoparticles growth. Powder XRD of all synthesised samples is available in Figure S2.

*2.2. Synthesis of SnO in alcohol-water mixtures*

To check the influence of water on the hydroxide-to-oxide transformation, various amounts of water (with an interval of 10 %) were added to the hydroxide dispersion in various alcohols. For methanol, 10 % water addition was sufficient for hydroxide-to-oxide transformation, leading to the formation of pure SnO layered nanoparticles. SEM images of the material formed in 70, 50, 30 and 10 % methanol content are presented in Figure 3, while for 80, 60, 40 and 20 % SEM images are available in ESI (Figure S3).

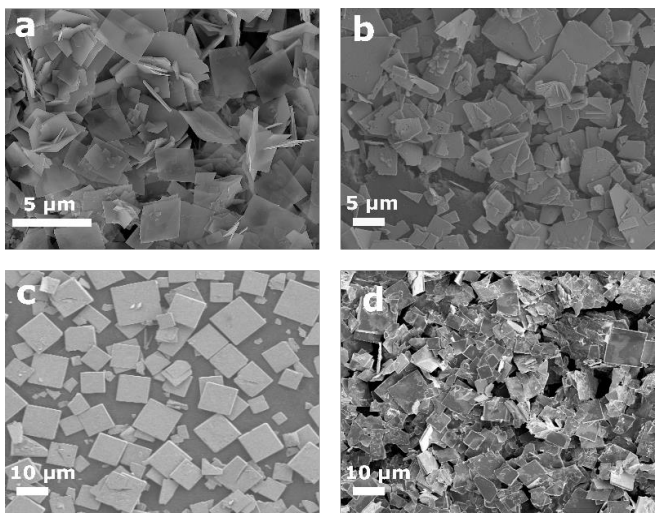

**Figure 3** SEM micrographs of the SnO samples produced in the methanol:water mixtures in the ratio a) 70:30, b) 50:50, c) 30:70 and d) 10:90.

The optimum conditions for the formation of regular thin SnO square-shape platelets were for methanol:water ratio of 80:20 and 70:30 (Figure S3a in ESI and 3a, respectively). When the methanol:water ratio was 60:40 and 50:50 (Figure S3b and 3b), nanoparticles were visibly crushed. Consequently, if the methanol content was further decreased to 40, 30, 20 and 10 % (Figure S3c, 3c, S3d and 3d), SnO particles became larger and thicker, which was a general trend observed for the methanol:water system, similar to results reported elsewhere.[23,24] Kumar *et al.* considered solvent polarity as the main factor affecting nucleation and growth of $SnO_2$ nanoparticles, and accordingly nanoparticles size and shape. Similar to our studies, they observed larger crystallite formation at higher water contents, which is likely related to the slower ionization and deposition in comparison to methanol heated sample. Comparable observations were reported for CdS,[18] ZnO,[21,25] Te[17] and ZnS.[26] The solvent-morphology relationship is also related to the nanoparticles surface energy and solvent-solvent interactions, and crystal growth is likely inhibited by the presence of alcohol molecules.[18] Water addition was also reported to influence SnO morphology, aggregation and stacking sequence.*[11,27]* Another interesting observation was reported for ZnS nanoparticles,[26] for which stronger interactions were both observed and calculated between water and ZnS nanoparticles than between ZnS and methanol. Zhang *et al.* focused on the surface environment rather than the particle size and reported reversible structural transformation upon water addition, which might be related to our studies.

In contrast to the methanol:water system, for other alcohols and their mixtures with water, no clear relationship was observed. One general observation was that with increasing water content, nanoparticles tend to grow thicker, but no consistent trend was observed. Interestingly, upon only 10 % water addition to ethanol and 2-propanol, the flower-like morphology was broken (Figure S4a, b). Further water addition (Figure S4c, d) resulted in the formation of thicker and square-like platelets, as observed for other alcohols. Similarly, flower-like hierarchical architectures composed of ordered microsheets reported by Sun *et al.* [12] were obtained via hydrothermal treatment, while the synthesis performed in pure alcohol was not described in that report.

For the reactions performed in 1-pentanol, and 1-hexanol, even the slightest amount of water dramatically changed the reaction kinetics; the colour change was observed within minutes, not hours as for the pure solvents. Hence, trace water was sufficient to induce immediate SnO formation, while long-chain alcohols significantly inhibited this process.[28] The bulky nature of the samples produced in more polar solvents lead to a lower number of active sites, due to lack of edges and corners. For this reason, for energy storage applications, we focused on the samples produced in methanol:water (70:30), ethanol, 1-hexanol and water, which represent various morphologies: platelets, flowers, thick squares and perforated thick squares, respectively.

*2.3. Detailed TEM characterization of SnO produced in methanol:water (70:30)*

The sample produced in the methanol:water (70:30) mixture was selected for additional TEM characterization, due to its thin two-dimensional nature and, consequently, predicted high electrochemical activity.

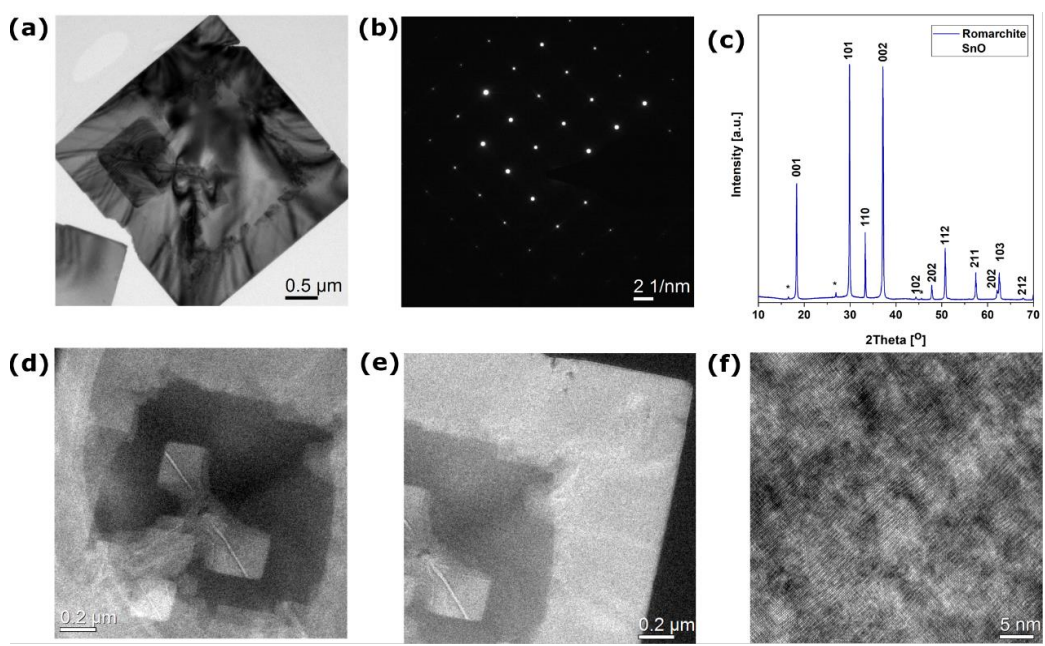

**Figure 4** TEM micrograph (a), SEAD (b), XRD (c), thickness map (d, e) and HRTEM (f) of the SnO sample produced in the methanol:water (70:30) mixture. Peaks marked with * are from the equipment.

TEM measurements (Figure 4a) confirmed that the obtained material consists of the square single crystal platelets with the side length of around 3 μm; particle size distribution is available in ESI (Figure S5). TEM analysis revealed that primarily formed platelets are crystallization centres for other crystals, which grow on top of those initially formed, shown in Figure 4a. Electron energy loss thickness maps presented in the Figures 4d and e additionally support this statement, where regions of higher intensity clearly indicate crystal formation on top of the existing ones. SEAD and XRD (Figure 4b and c, respectively) proved high crystallinity, confirming single Romarchite phase SnO (JCPDS 06-0395 space group P4/nmm), without any impurities. HRTEM analysis (Figure 4f) provides additional evidence of the high quality of the produced material. Exfoliation of these layered nanostructures will be considered for future work.

Similar square nanoplatelets were produced by Wang *et al.*[14] by the dissolution-recrystallization. Their synthesis method relied on the formation of tin oleate complex and then injection of tri-n-octylamine (TOA) at 250 °C under $N_2$ atmosphere and subsequent heating to 340 °C. In the reported experiment, TOA played a double role, firstly activating the tin oleate complex decomposition and secondly acting as a surfactant, which controlled the shape of the SnO nanoparticles. It is worth highlighting here that protective $N_2$ atmosphere was crucial, as under air intermediate tin oleate complex decomposed directly to $SnO_2$. A similar mechanism likely took place in our studies, where intermediate tin oxide-hydroxide was decomposing to tin(II) oxide while heated in methanol:water mixture 70:30 ($B_p$ of 78 °C), and the solvent particles selectively adsorbed onto the (001) plane of the SnO and supressed its growth rate, leading to the formation of square platelets.[29]

*2.4. Synthesis in water and theoretically predicted SnO nanoparticle shape*

To further augment our understanding of the underlying chemical mechanisms determining the resultant nanoparticle morphology, DFT calculations were performed including dispersion corrections. Structural relaxation of SnO predicted equilibrium lattice parameters of *a,b* = 3.83 Å, *c* = 4.79 Å, in good agreement with experiment (less than 1% deviation) (Table S1 in ESI).[30]

While the morphology of solution-grown crystals is dependent on many factors, including solvent properties, precursor concentrations, nucleation kinetics etc., the equilibrium crystal shape is primarily determined by the surface energetics.[31,32] This dependence of the crystal shape on surface energetics is embodied by the Gibbs-Wulff theorem, which states that the shape of a crystalline material is given by the polyhedron that minimises the overall surface energy. Although large single-crystal materials rarely attain their true equilibrium shape, unless grown under specialised conditions, nanoparticles can often reach equilibrium within seconds due to their small size. This allows the prediction of nanoparticle morphology via a Wulff construction, given knowledge of the corresponding surface energies of crystal facets. The calculated surface energies of each unique low-index SnO crystal facet, up to a maximum Miller index of 2, are provided in Table S2 in the ESI, with the corresponding Wulff shape of SnO shown in Figure 5.

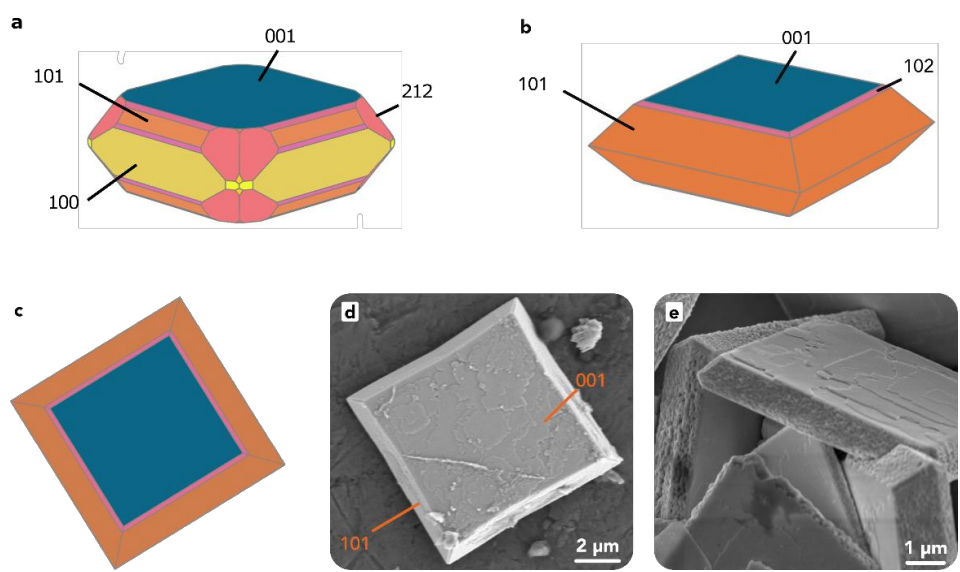

**Figure 5** Predicted Wulff equilibrium crystal shape for SnO, including all low-index surfaces (a) and only allowing crystal faces with surface energies less than 0.6 J/m$^2$ (b, c), alongside SEM micrographs of experimentally-synthesised platelets in water (d, e).

The predicted crystal shape is in good agreement with that observed for the SnO platelets grown in pure H$_2$O solvent (Figure 5d, e and S1 in ESI), as well as in 1-butanol, 1-pentanol and 1-hexanol (Figure 2d-f). The layered structure of SnO begets an extremely low surface energy of 0.25 J/m$^2$ for the {001}[33] facet, due to the absence of cleaved bonds, therefore dominating the minimum-energy crystal shape. The result is a square-like crystal platelet, as shown in Figure 5. The surface-weighted ionisation potential, electron affinity and work function for this crystal morphology were calculated as 4.47, 3.79 and 4.25 eV respectively, in excellent agreement with experiment (see ESI for further details).[34] The primary difference between the calculated Wulff shape (including all low-index surfaces) (Figure 5a) and the experimentally-observed platelet morphology (for the SnO samples prepared in pure $H_2O$, Figure 5d, e and longer chain alcohols Figure 2d-f) is the absence of the {100} and {212} faces in the observed platelet shape.

A possible origin of the minor discrepancy is that the {001} and {101} surfaces are stabilised by the highly-polar water solvent, relative to the {100} and {212} crystal faces, exposing greater densities of steps and kinks. Indeed, the {001} and {101} surfaces exhibit an adsorbate-binding-site density of 0.068 and 0.085 Sites/$Å^2$, for $Sn^{2+}$ within 1 Å of the surface (where metal-hydroxyl bonding would occur), compared to 0.040 and 0.054 Sites/$Å^2$ for the {212} and {100} faces, respectively. This allows for stronger solvent-platelet interactions at the {001} and {101} crystal faces, further lowering the surface energies and favouring their dominance in the resulting crystal shape.[32]

Another possible reason for the absence of the {100} and {212} faces is that the platelets may not have reached the "thermodynamic limit" of large particles.[32,35] In this case, the formation of the higher energy surfaces – {100} and {212} with $E_{surface}$ = 0.61 and 0.64 J/$m^2$ respectively, relative to 0.57 J/$m^2$ for the {101} surface – may be kinetically-inhibited, despite being thermodynamically-favourable due to the reduction in particle surface area. In either of these cases, the dominance of the {001} and {101} faces would lead to the crystal shape shown in Figure 5b,c, which is in good agreement with the experimental SEM results (Figures 5d,e, 2d-f and S1 in ESI).

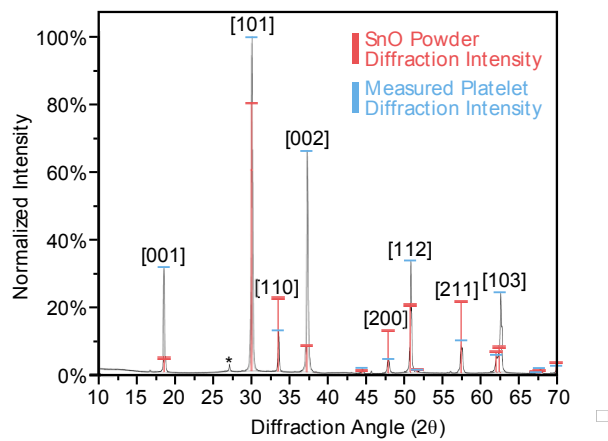

**Figure 6** Labelled X-ray diffraction pattern of synthesised SnO platelets, alongside comparison to typical relative peak intensities for SnO powder (normalised by the integral signal intensity of the sample). The normalised intensity is given with respect to the [101] peak. The asterisk indicates an equipment peak.

From this analysis, alongside inspection of the XRD (Figure 6) and SEM results (Figures 2,3,5) and TEM results (Figure 4) we may assign the observed platelet surfaces as {001} and {101}, respectively, as depicted in Figure 5d. Indeed, this assignment is supported by the XRD data (Figure 6), which exhibits significantly greater peak intensities for the {001}, {002}, {101} and {103} crystallographic directions, and reduced intensities for the {110}, {200} and {211} directions, relative to the diffractogram of SnO powder (JCPDS No. 06-0395). This indicates a preferential orientation of crystallites within the platelets along these crystal directions, which often occurs when these crystal planes align with the Miller index of the particle surface.[13,36-38]

From the significant variation in crystal morphology observed for different growth conditions, it is evident that solvent properties strongly influence both the kinetics and thermodynamics of the crystal growth in solution, thus producing crystal platelets which deviate from the *in vacuo* equilibrium shape in certain solvents (such as EtOH) and their mixtures with water.

*2.5. SnO particles morphology influence on the energy storage*

To investigate the energy storage capabilities of the different morphologies of SnO, lithium-ion battery electrodes were produced from thick squares (SnO produced in $H_2O$), flowers (ethanol), platelets (methanol:water, 70:30) and perforated thick squares (1-hexanol). In each case, SnO material was combined with 15 % mass fraction of P3-single walled carbon nanotubes (SWCNT) to form composite electrodes. While SWCNTs are not themselves suited toward energy storage applications, their incorporation allows them to act as a conductive additive at a lower weight loading than other conventional carbon additives (carbon black/graphite), presenting a more effective strategy to form an effective electrical percolation network.[39] In addition CNTs remove the need to incorporate a polymer binders in the composite further improving overall battery performance.

All SnO tested from the various solvent conditions produced high capacity anodes. The impact of particle morphology on electrochemical performance is visible upon inspection of the rate capability (Figure 7a) of the various SnO structures. The thinner, platelet-like morphologies of SnO formed in ethanol and 70% methanol produced the highest capacities at low C-rates, with initial discharges of 960 mAh.g$^{-1}$ and 985 mAh.g$^{-1}$ respectively far above the theoretical capacity of 875 mAh.g$^{-1}$ (note initial discharge is not the first-discharge as this would include the irreversible capacity of the material attributed to the $LiO_2$ and interphase formation).[5] The reduced dimensions and significantly-increased surface area of the platelet-like structures permit greater contact between electrode and electrolyte resulting in higher $Li^+$ ion flux across the interface and thus greater charge capacity. The relatively bulkier morphologies created by the synthesis involving $H_2O$ and 1-hexanol produce anodes that do not meet the theoretical capacity of SnO.

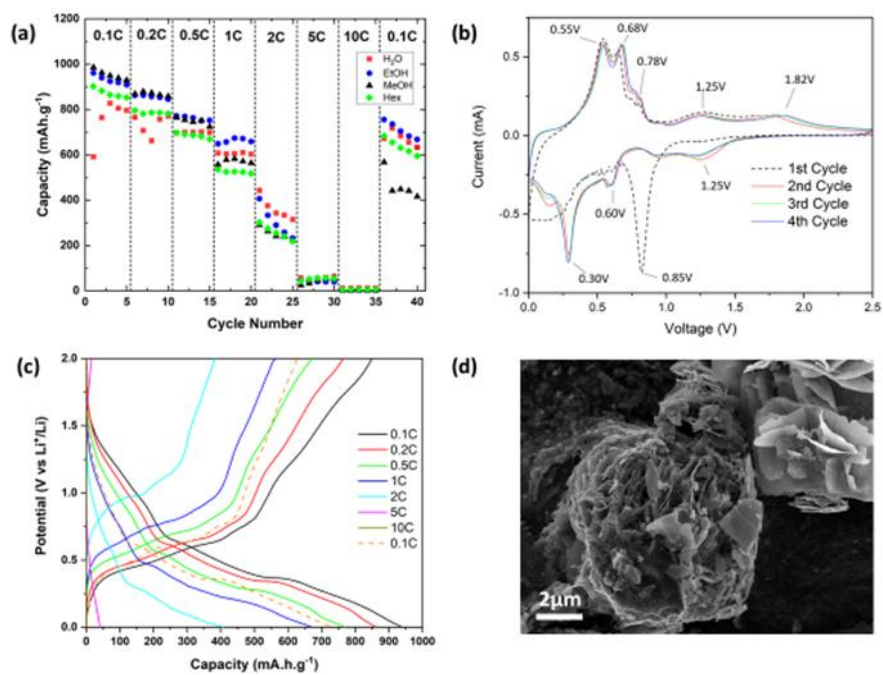

**Figure 7** Electrochemical properties of SnO composites: (a) rate capability of SnO formed in H$_2$O, ethanol, methanol:water(70:30) and hexanol (b) cyclic voltammograms acquired between 0 V and 2.5 V of SnO formed in ethanol at a scan rate of 0.1 mV/s (c) representative charge-discharge curve of SnO formed in ethanol (d) SEM image of SnO/P3-SWCNT(15%) composite

The alloying/dealloying mechanisms of the material produced in ethanol were evaluated by cyclic voltammetry at a scan rate of 0.1 mV/s. During the initial cathodic sweep, the large irreversible peak at 0.85 V indicates the irreversible formation of lithium dioxide and metallic tin.[33,40]. This is an attractive attribute of transition metal oxide anodes as the inactive-matrix composite of Li$_2$O plays a role in accommodating the large volumetric changes the anode will be subject to upon subsequent charge/discharges. The metallic Sn can be further lithiated into Li$_x$Sn ($0 \leq x \leq 4.4$) and this is seen at the reductive peaks at 0.60 and 0.30 V. The peaks at 0.55, 0.68 and 0.78 V in the anodic sweep correspond to the delithiation of the various Li-Sn alloys formed during the charge process, whilst the broad peak at 1.25 V is due to the formation of the pure Sn.[41]

The enhanced stability of the nanoflower morphology (EtOH) is shown by its capacity retention of 79% (Figure 7c) upon recycling at 0.1C relative to the platelet (MeOH:H$_2$O) retention of 60% (Figure S10c). The contacted points between the platelets in the nanoflowers may provide additional paths for electron flow perpendicular to the plane, and this would account for the enhanced stability over multiple cycles.[42] The large capacity of alloying anodes is accompanied by the massive volumetric change in the electrode (360% in the Li-Sn system[43]) and this leads to the pulverization of the active material during repeated cycling and capacity fade. Upon cycling the nanostructured SnO anodes at harsh conditions of 1C, all anodes maintain a capacity greater than 400 mAh.g$^{-1}$ for the first 50 cycles, greater than that of the theoretical capacity of the commercial graphite anode of 372 mAh.g$^{-1}$ (Figure S11).[44] Further electrochemical characterization of produced SnO nanostructures will be performed.

3. **Conclusions**

In summary, we produced tin(II) oxide nanoparticles with various morphologies by refluxing tin(II) hydroxide in numerous alcohols (methanol to 1-hexanol) and their mixtures with water. We observed that nanoparticles size and shape depend strongly on the solvent nature, and this morphology-solvent relationship is very complex, although polarity and reaction temperature seem to be key factors. Most probably, particular solvents (methanol, ethanol) adsorb selectively on the specific crystal planes directing crystal growth in certain directions. Moreover, we predicted the thermodynamically-favoured crystal shape (Wulff construction) which was in good agreement to the sample produced in water as well as 1-butanol, 1-pentanol and 1-hexanol. Future theoretical work focusing on the solvent - crystal surface interaction, and hence the thermodynamics and kinetics of crystal growth in solution, would permit further understanding of the complex relationship between solvent properties and crystal morphology. Finally, we compared nanoparticles with different morphology in energy storage application and we observed significant differences in their performance. Capacities at low C-rates, with initial discharges of 960 mAh.g$^{-1}$ and 985 mAh.g$^{-1}$ were produced for ethanol and 70 % methanol, respectively, while samples with bulkier morphologies delivered significantly lower values.

## 4. Methods

**Chemical materials:** anhydrous tin(II) chloride (Alfa Aesar; 98%), ammonia solution (Merck, 25% w/w), anhydrous ethanol (Merck; ≤99.5%), methanol (Merck; 99.8%), 1-propanol (Merck; 99.7%), 1-butanol (Merck; 99.8%), 1-pentanol (Sigma Aldrich, ≥ 99%), 1-hexanol (Alfa Aesar, 99%) and ethylene glycol (Merck; 99.8%) were used as provided by the manufacturer, with no additional purification.

**Synthesis method:** Tin(II) oxide nanoplatelets were produced by a two-step wet-chemistry method, involving firstly precipitation of tin(II) hydroxide and its subsequent thermal transformation to tin(II) oxide, by refluxing in a solvent media.

Preparation of tin(II) hydroxide: A 0.1 M tin(II) chloride solution was prepared by dissolving 7.584 g (0.04 mol) of $SnCl_2$ in 400 ml of ice-cold deionised water. Then, 12.8 ml (0.08 mol) of ammonia was added dropwise to the $SnCl_2$ solution with simultaneous stirring. The white precipitate obtained was washed with deionised water and ethanol. Tin(II) hydroxide was freshly prepared before each synthesis to avoid its degradation.

In the second step, 3.000 g of tin(II) hydroxide obtained in the first synthesis step was dispersed in 60 ml of the solvent under study and refluxed for 24 hours. After heating, the sample was washed with deionized water and ethanol. Materials were stored as dispersions in ethanol at room temperature.

**Characterization techniques:** Scanning electron microscopy (SEM) analysis was performed using Zeiss Ultra Plus (Carl Zeiss AG, Germany) operating at 3-5 kV. Transmission electron microscopy (TEM) images, selected area electron diffraction (SAED) and high-resolution analysis were acquired using FEI Titan (FEI, Oregon, USA) microscope operated at 300 keV. Corresponding electron energy loss thickness maps were acquired using a Gatan Imaging Filter (Gatan, USA). Powder X-Ray diffraction was measured using a Bruker Advance Powder X-ray diffractometer equipped with a Mo-Kα emission source (λ=0.7107 Å) in the Bragg-Brentano configuration.

**Electronic structure theory:** All calculations were performed using Density Functional Theory within periodic boundary conditions through the Vienna Ab Initio Simulation Package (VASP).[45-48]

The optB86b-vdW[49] Density Functional Theory exchange-correlation functional was used for geometry optimizations and calculations of surface energies. This choice was motivated by the ability of this dispersion-corrected functional to accurately incorporate van der Waals interactions in solids, yielding good agreement with experimental measurements of interlayer spacings (Table S1 in ESI) and binding energies (in layered solids).[50-52] Scalar-relativistic pseudopotentials were employed, and the projector-augmented wave method was used to describe the interaction between core and valence electrons.[53]

To obtain an accurate description of electronic properties, namely the absolute band positions, Hybrid DFT was implemented. Specifically, a modified PBE0 Hybrid DFT functional, with 17% exact Hartree-Fock exchange, was employed, giving close agreement with reported values for the direct and indirect band gaps (Figure S6-S9 in ESI).[54] Vacuum alignment of the electrostatic potential offset $\Delta V$ was performed via the method of Butler *et al*,[55] using the cheaper optB86b-vdW[49] DFT functional, due to the vastly-reduced computational expense and well-established accuracy for this form of calculation (further details are available in ESI).[56,57]

A convergence criterion of 0.005 eV/Å was imposed on the individual atomic forces during structural optimization. Structural relaxation and ground-state energy calculations were carried out using a $6 \times 6 \times 4$ Gamma-centred Monkhorst-Pack *k*-point mesh, with a well-converged 850 eV plane-wave kinetic energy cutoff. In each calculation, convergence with respect to k-point density and plane-wave energy cutoff was confirmed for the property of interest. Additionally, calculated surface energies were converged with respect to supercell slab and vacuum sizes, for which thicknesses of 10 Å each gave convergence to within 0.02 J/m$^2$. The equilibrium structures and input calculation parameters for the crystal materials investigated, alongside convergence analysis, are provided in an online repository at doi.org/10.5281/zenodo.4030515

**Electrochemical measurements:** Electrodes were prepared by mixing 85% of active material with 15% P3-SWCNT in IPA. This dispersion was filtered under vacuum onto a polyolefin membrane (15μm) and electrodes were cut using an electrode puncher (EL-CELL, 12mm). EL-CELL were used to assemble the coin cells in an Argon filled glove box (MBRAUN) with lithium foil as a counter electrode, a polyolefin separator and 1M LiFP$_6$ in EC:DMC (50:50) as electrolyte. Cyclic voltammetry and galvanostatic charge-discharge measurements were performed in a BioLogic VMP 300 and analyzed using the EC-Lab software.

**Data availability**

The data related to the results reported in this work are available from the corresponding authors, subject to reasonable request.

**Acknowledgements**

Sonia Jaśkaniec and Valeria Nicolosi would like to thank the following funding support: Science Foundation Ireland (AMBER) and European Research Council (CoG and 3D2DPrint).
 Seán R. Kavanagh acknowledges the EPSRC Centre for Doctoral Training in the Advanced Characterisation of Materials (CDT-ACM)(EP/S023259/1) for funding a PhD studentship and the Imperial College Research Computing Service ([doi.org/10.14469/hpc/2232](doi.org/10.14469/hpc/2232)) for computational resources. Via membership of the UK's HEC Materials Chemistry Consortium, which is funded by the EPSRC (EP/L000202, EP/R029431), this work used the ARCHER UK National Supercomputing Service ([www.archer.ac.uk](www.archer.ac.uk)) and the UK Materials and Molecular Modelling (MMM) Hub (Thomas). DOS and AW acknowledge funding from the Faraday Institution (FIRG003).

**Author contributions**

S.J. and V.N. designed the project and discussed the results. S.J. planned the synthesis procedure and together with S.R.K. performed chemical reactions, SEM and XRD analyses. S.R.K designed and performed DFT calculations. J.C. and S.R. completed electrochemistry measurements. C.H. performed TEM characterization. A.W. and D.O.S. discussed DFT results. S.J. structured and wrote the paper with all authors contributing to the manuscript.

**Additional information**

**Supplementary information** available online on the npj 2D Materials and Applications website.

**Competing interests:** The authors declare no competing interests.

**References**

1. Suresh, S. J. N. N. Semiconductor nanomaterials, methods and applications: a review. **3**, 62-74 (2013).
2. Jun, Y. w., Choi, J. s. & Cheon, J. J. A. C. I. E. Shape control of semiconductor and metal oxide nanocrystals through nonhydrolytic colloidal routes. **45**, 3414-3439 (2006).
3. Wu, Z., Yang, S. & Wu, W. J. N. Shape control of inorganic nanoparticles from solution. **8**, 1237-1259 (2016).
4. Coelho, J. *et al.* Manganese oxide nanosheets and a 2D hybrid of graphene–manganese oxide nanosheets synthesized by liquid-phase exfoliation. *2D Materials* **2**, 025005 (2015).
5. Zhang, F., Zhu, J., Zhang, D., Schwingenschlögl, U. & Alshareef, H. N. J. N. l. Two-dimensional SnO anodes with a tunable number of atomic layers for sodium ion batteries. **17**, 1302-1311 (2017).
6. Walsh, A. & Watson, G. W. Electronic structures of rocksalt, litharge, and herzenbergite SnO by density functional theory. *Physical Review B* **70**, 235114 (2004).
7. Slater, M. D., Kim, D., Lee, E. & Johnson, C. S. Sodium-ion batteries. *Advanced Functional Materials* **23**, 947-958 (2013).
8. Idota, Y., Kubota, T., Matsufuji, A., Maekawa, Y. & Miyasaka, T. Tin-based amorphous oxide: a high-capacity lithium-ion-storage material. *Science* **276**, 1395-1397 (1997).
9. Iqbal, M. Z. *et al.* Structural and electrochemical properties of SnO nanoflowers as an anode material for lithium ion batteries. **67**, 665-668 (2012).
10. Song, H. *et al.* Controllable lithium storage performance of tin oxide anodes with various particle sizes. *international journal of hydrogen energy* **40**, 14314-14321 (2015).
11. Ning, J. *et al.* Syntheses, characterizations, and applications in lithium ion batteries of hierarchical SnO nanocrystals. **113**, 14140-14144 (2009).
12. Sun, Y.-H., Dong, P.-P., Lang, X. & Nan, J.-M. J. C. C. L. A novel rose flower-like SnO hierarchical structure synthesized by a hydrothermal method in an ethanol/water system. **25**, 915-918 (2014).
13. Uchiyama, H., Ohgi, H., Imai, H. J. C. g. & design. Selective preparation of SnO2 and SnO crystals with controlled morphologies in an aqueous solution system. **6**, 2186-2190 (2006).
14. Wang, S.-C., Chiang, R.-K. & Hu, P.-J. J. J. o. t. E. C. S. Morphological and phase control of tin oxide single-crystals synthesized by dissolution and recrystallization of bulk SnO powders. **31**, 2447-2451 (2011).
15. Liu, W. *et al.* One-step synthesis of SnO hierarchical architectures under room temperature and their photocatalytic properties. **29**, 284002 (2018).
16. Wang, S. *et al.* Solution route to single crystalline SnO platelets with tunable shapes. 507-509 (2005).
17. Liu, J. *et al.* Understanding the solvent molecules induced spontaneous growth of uncapped tellurium nanoparticles. **6**, 32631 (2016).
18. Thakur, P. & Joshi, S. S. J. J. o. E. N. Effect of alcohol and alcohol/water mixtures on crystalline structure of CdS nanoparticles. **7**, 547-558 (2012).
19. Ávila, H. & Rodríguez-Páez, J. J. J. o. N.-C. S. Solvent effects in the synthesis process of tin oxide. **355**, 885-890 (2009).
20. Ejaz, A., Jeon, S. J. B. & Bioelectronics. The insight study of SnO pico size particles in an ethanol-water system followed by its biosensing application. (2018).

21  Ungula, J. & Dejene, B. J. P. B. C. M. Effect of solvent medium on the structural, morphological and optical properties of ZnO nanoparticles synthesized by the sol–gel method. **480**, 26-30 (2016).
22  Han, Z. *et al.* Solvothermal preparation and morphological evolution of stannous oxide powders. **48**, 99-103 (2001).
23  Jiang, L. *et al.* Size-controllable synthesis of monodispersed SnO2 nanoparticles and application in electrocatalysts. **109**, 8774-8778 (2005).
24  Kumar, V. *et al.* Effect of solvent on crystallographic, morphological and optical properties of SnO2 nanoparticles. **85**, 202-208 (2017).
25  Uekawa, N., Kitamura, M., Ishii, S., Kojima, T. & Kakegawa, K. Low-temperature synthesis of ZnO nanoparticles by heating of Zn (OH) 2 in a neutral mixed solution of ethanol and H2O. *Nippon seramikkusu kyokai gakujutsu ronbunshi* **113**, 439-441 (2005).
26  Zhang, H., Gilbert, B., Huang, F. & Banfield, J. F. Water-driven structure transformation in nanoparticles at room temperature. *Nature* **424**, 1025-1029 (2003).
27  Burrows, N. D. *et al.* Crystalline nanoparticle aggregation in non-aqueous solvents. *CrystEngComm* **16**, 1472-1481 (2014).
28  Zherebetskyy, D. *et al.* Hydroxylation of the surface of PbS nanocrystals passivated with oleic acid. *Science* **344**, 1380-1384, doi:10.1126/science.1252727 (2014).
29  Wang, Z.    (ACS Publications, 2000).
30  Pannetier, J. & Denes, G. Tin (II) oxide: structure refinement and thermal expansion. *Acta Crystallographica Section B: Structural Crystallography and Crystal Chemistry* **36**, 2763-2765 (1980).
31  Wulff, G. Xxv. zur frage der geschwindigkeit des wachsthums und der auflösung der krystallflächen. *Zeitschrift für Kristallographie-Crystalline Materials* **34**, 449-530 (1901).
32  Barmparis, G. D., Lodziana, Z., Lopez, N. & Remediakis, I. N. Nanoparticle shapes by using Wulff constructions and first-principles calculations. *Beilstein journal of nanotechnology* **6**, 361-368 (2015).
33  Huggins, R. *Advanced batteries: materials science aspects*.  (Springer Science & Business Media, 2008).
34  Li, X. *et al.* Determination of some basic physical parameters of SnO based on SnO/Si pn heterojunctions. *Applied Physics Letters* **106**, 132102 (2015).
35  Ringe, E., Van Duyne, R. P. & Marks, L. D. Kinetic and thermodynamic modified Wulff constructions for twinned nanoparticles. *The Journal of Physical Chemistry C* **117**, 15859-15870 (2013).
36  Singh, M. *et al.* Soft exfoliation of 2D SnO with size-dependent optical properties. *2D Materials* **4**, 025110 (2017).
37  Liang, L., Sun, Y., Lei, F., Gao, S. & Xie, Y. J. J. o. M. C. A. Free-floating ultrathin tin monoxide sheets for solar-driven photoelectrochemical water splitting. **2**, 10647-10653 (2014).
38  Okamura, K., Nasr, B., Brand, R. A. & Hahn, H. Solution-processed oxide semiconductor SnO in p-channel thin-film transistors. *Journal of Materials Chemistry* **22**, 4607-4610 (2012).
39  Landi, B. J., Ganter, M. J., Cress, C. D., DiLeo, R. A. & Raffaelle, R. P. Carbon nanotubes for lithium ion batteries. *Energy & Environmental Science* **2**, 638-654 (2009).
40  Shin, J. H. & Song, J. Y. Electrochemical properties of Sn-decorated SnO nanobranches as an anode of Li-ion battery. *Nano Convergence* **3**, 1-7 (2016).
41  Mohamedi, M. *et al.* Amorphous tin oxide films: preparation and characterization as an anode active material for lithium ion batteries. *Electrochimica Acta* **46**, 1161-1168 (2001).
42  Wang, B. *et al.* Folding graphene film yields high areal energy storage in lithium-ion batteries. *ACS nano* **12**, 1739-1746 (2018).

43  Tamura, N., Ohshita, R., Fujimoto, M., Kamino, M. & Fujitani, S. Advanced structures in electrodeposited tin base negative electrodes for lithium secondary batteries. *Journal of The Electrochemical Society* **150**, A679 (2003).
44  Goriparti, S. *et al.* Review on recent progress of nanostructured anode materials for Li-ion batteries. *Journal of power sources* **257**, 421-443 (2014).
45  Kresse, G. & Hafner, J. Norm-conserving and ultrasoft pseudopotentials for first-row and transition elements. *Journal of Physics: Condensed Matter* **6**, 8245 (1994).
46  Kresse, G. & Hafner, J. Ab initio molecular dynamics for liquid metals. *Physical Review B* **47**, 558 (1993).
47  Kresse, G. & Furthmüller, J. Efficiency of ab-initio total energy calculations for metals and semiconductors using a plane-wave basis set. *Computational materials science* **6**, 15-50 (1996).
48  Kresse, G. & Furthmüller, J. Efficient iterative schemes for ab initio total-energy calculations using a plane-wave basis set. *Physical review B* **54**, 11169 (1996).
49  Klimeš, J., Bowler, D. R. & Michaelides, A. Van der Waals density functionals applied to solids. *Physical Review B* **83**, 195131 (2011).
50  Lozano, A., Escribano, B., Akhmatskaya, E. & Carrasco, J. Assessment of van der Waals inclusive density functional theory methods for layered electroactive materials. *Physical Chemistry Chemical Physics* **19**, 10133-10139 (2017).
51  Tawfik, S. A., Gould, T., Stampfl, C. & Ford, M. J. Evaluation of van der Waals density functionals for layered materials. *Physical Review Materials* **2**, 034005 (2018).
52  Tran, F., Kalantari, L., Traoré, B., Rocquefelte, X. & Blaha, P. Nonlocal van der Waals functionals for solids: Choosing an appropriate one. *Physical Review Materials* **3**, 063602 (2019).
53  Blöchl, P. E. Projector augmented-wave method. *Physical review B* **50**, 17953 (1994).
54  Allen, J. P., Scanlon, D. O., Piper, L. F. & Watson, G. W. Understanding the defect chemistry of tin monoxide. *Journal of Materials Chemistry C* **1**, 8194-8208 (2013).
55  Butler, K. T., Hendon, C. H. & Walsh, A. Electronic chemical potentials of porous metal–organic frameworks. *Journal of the American Chemical Society* **136**, 2703-2706 (2014).
56  Li, Y.-H. *et al.* Revised ab initio natural band offsets of all group IV, II-VI, and III-V semiconductors. *Applied Physics Letters* **94**, 212109 (2009).
57  Li, Z. *et al.* Bandgap Lowering in Mixed Alloys of Cs2Ag (SbxBi1-x) Br6 Double Perovskite Thin Films. *arXiv preprint arXiv:2007.00388* (2020).